\documentclass[sigconf,nonacm,9pt]{acmart}

\AtBeginDocument{%
  }

 \copyrightyear{2026}
\acmYear{2026}
\setcopyright{cc}
\setcctype{by}
\acmConference[ICCAD '26]{IEEE/ACM International Conference on Computer-Aided Design}{November 08--12, 2026}{San Jose, CA, USA}
\acmBooktitle{IEEE/ACM International Conference on Computer-Aided Design (ICCAD '26), November 08--12, 2026, San Jose, CA, USA}
\acmDOI{10.1145/3831252.3834153}
\acmISBN{979-8-4007-2873-0/2026/11}

\usepackage{comment}
\usepackage{qcircuit}
\usepackage{xypic}
\usepackage{url}

\begin{document}

\title{SQD-Enabled Circuit Compression for Resource-Efficient Quantum Chemistry}

 \author{Kangyu Zheng}
 \email{kyzheng@cse.cuhk.edu.hk}
 \affiliation{%
   \institution{The Chinese University of Hong Kong}
   \city{Hong Kong}
   \country{China}
 }

 \author{Yidong Zhou}
 \email{yidong.zhou@rutgers.edu}
 \affiliation{%
   \institution{Rutgers University}
   \city{New Brunswick}
   \state{New Jersey}
   \country{USA}
 }
 
 \author{Jinglei Cheng}
 \email{jinglei@eigenai.com}
 \affiliation{%
   \institution{Eigen AI}
   \city{Palo Alto}
   \state{California}
   \country{USA}
 }

 \author{Zhemin Zhang}
 \email{hisenzhang01@gmail.com}
 \affiliation{%
   \institution{The Chinese University of Hong Kong}
   \city{Hong Kong}
   \country{China}
 }

 \author{Shaohua Li}
 \email{shaohuali@cuhk.edu.hk}
 \affiliation{%
   \institution{The Chinese University of Hong Kong}
   \city{Hong Kong}
   \country{China}
 }

 \author{Zhiding Liang}
 \email{zliang@cse.cuhk.edu.hk}
 \affiliation{%
   \institution{The Chinese University of Hong Kong}
   \city{Hong Kong}
   \country{China}
 }







\renewcommand{\shortauthors}{Trovato et al.}

\begin{abstract}
Sample-based Quantum Diagonalization (SQD) recovers ground-state energies by classically diagonalizing a Hamiltonian in the subspace spanned by quantum samples, requiring only bitstrings with sufficient ground-state overlap rather than an accurate variational energy. We reveal and exploit this underexplored robustness property: how much non-Clifford and variational expressivity can be removed from the sampling circuit before SQD accuracy degrades? We answer through two complementary compression techniques: gradient-based operator pruning, which discards low-impact excitation operators, and Clifford rounding, which snaps remaining parameters to the nearest Clifford angle. Both of these techniques can be applied to a VQE ansatz on a qubit-reduced Hamiltonian. A systematic ablation study across 21 molecules shows that median SQD error stays within chemical accuracy even at 50\% compression on both axes, while simulation speedup reaches $33\times$. Hardware validation on 6 molecules on IBM quantum hardware confirms up to $2.8\times$ transpiled-depth reduction with zero loss in SQD accuracy. Our implementation can be found at: \url{https://github.com/zkysfls/cs-vqe-sqd}
\end{abstract}



\received[accepted]{11 July 2026}

\maketitle

\section{Introduction}
\label{sec:intro}
Quantum computing (QC) presents a fundamentally different computational approach capable of addressing classically intractable problems, with applications in many areas~\cite{shor1999polynomial, kandala2017hardware, moll2018quantum,biamonte2017quantum}. The current Noisy Intermediate-Scale Quantum (NISQ) era is characterized by systems with several hundred qubits that exhibit significant noise limitations~\cite{preskill2018quantum}. Despite advances in error mitigation techniques to improve operational fidelity~\cite{barron2020measurement, botelho2022error, czarnik2021error, ding2020systematic, liao2024machine}, performance remains inadequate for most practical applications~\cite{quek2024exponentially}.

Variational Quantum Algorithms (VQAs) are well-suited for NISQ hardware implementation. These hybrid quantum-classical methods have shown utility across multiple domains, from molecular simulations~\cite{peruzzo2014variational, grimsley2019adaptive, zhou2024quantum, zheng2025qcsadmequantumcircuitsearch} to optimization problems~\cite{moll2018quantum}. VQAs utilize parameterized quantum circuits whose variables are optimized using classical computational methods to minimize problem-specific objective functions. While their adaptability to hardware characteristics is advantageous, their performance on current quantum processors often fails to meet precision requirements in fields such as quantum chemistry, especially at larger scales~\cite{wang2024can, kandala2017hardware}.

VQAs face several persistent challenges, including optimization landscapes with barren plateaus, vulnerability to quantum hardware noise, limitations in implementable circuit depths, and difficulties in the optimal allocation of tasks between classical and quantum computational resources~\cite{ravi2022cafqa, 10.1145/3622781.3674178}. These challenges intensify with increasing problem size and when addressing applications with high precision requirements.

To address these challenges, we propose a resource-efficient hybrid pipeline that reduces qubit count and circuit depth while preserving chemical accuracy. Our approach begins with a qubit-reduction step (Contextual-Subspace~\cite{Weaving2025} in this work, though other methods may serve the same role): the Hamiltonian is split into a noncontextual component that is solved classically and a smaller contextual remainder targeted by VQE on hardware. This partitioning shrinks the qubit register and measurement load without discarding strongly correlated physics.

After pre-processing reduction, our pipeline constructs a problem-specific ansatz by performing a classical gradient analysis like ADAPT-VQE~\cite{grimsley2019adaptive}. This allows us to select only the most significant excitation operators relevant to the reduced Hamiltonian, resulting in a low-depth, compact ansatz with few parameters. We further compress the circuit by rounding the parameters of the least significant operators to the nearest Clifford angle, converting costly non-Clifford gates into classically trackable Clifford operations.

A key observation motivating these aggressive simplifications is that SQD~\cite{Robledo_Moreno_2025} does not require an accurate VQE energy, it only requires the circuit to produce bitstrings with sufficient overlap on the ground-state subspace. SQD then forms low-dimensional subspaces from these samples and classically diagonalizes the Hamiltonian to refine energies and eigenstates. This decoupling of circuit quality from final accuracy creates room for circuit compression that would be unacceptable in a standalone VQE.

In summary, our main contributions are:
\begin{itemize}
    \item The identification that SQD's tolerance to circuit imperfection enables aggressive circuit compression---gradient pruning and Clifford rounding---without degrading final energy accuracy.
    \item We present a systematic ablation study across 21 molecules characterizing the two-dimensional accuracy: depth tradeoff of gradient pruning and Clifford rounding, demonstrating that the median error remains within chemical accuracy even at 50\% compression along both axes.
    \item Hardware validation on 6 molecules on IBM quantum hardware, demonstrating up to $2.8\times$ reduction in transpiled circuit depth with zero loss in SQD accuracy.
\end{itemize}

\section{Background}
\label{sec:background}
\subsection{Variational Quantum Algorithms and Classical Simulation}
VQAs represent a leading approach for near-term quantum processors, employing a hybrid quantum-classical strategy~\cite{Cerezo_2021}. A parameterized quantum circuit (the \textbf{Ansatz}) is executed on quantum hardware while a classical optimizer trains the parameters $\boldsymbol{\theta}$ to minimize a \textbf{Cost Function} $C(\boldsymbol{\theta})$. Variational Quantum Eigensolver (VQE)~\cite{peruzzo2014variational}, one of the most prominent VQA variants, finds ground-state energies through minimization of the expectation value $C(\boldsymbol{\theta}) = \langle \psi(\theta) | H | \psi(\theta) \rangle$.

A fundamental challenge in VQE is balancing classical tractability with quantum expressivity~\cite{sim2019expressibility, holmes2022connecting}. Clifford-assisted methods~\cite{ravi2022cafqa, 10.1145/3622781.3674178} restrict ansätze to Clifford gates for guaranteed classical simulability, but this severely limits expressivity for capturing complex correlations. Recent optimization approaches~\cite{li2024efficient, liang2024napa} focus on reducing energy evaluations through sequential optimization or pulse-level refinement, yet treat parameter optimization as independent of ansatz design without addressing the underlying expressivity-tractability trade-off~\cite{arrasmith2022equivalence}.

Rather than fixing an ansatz structure a priori, adaptive ansatz selection methods~\cite{rudolph2023synergistic, tang2021qubit} improve expressivity efficiency by dynamically constructing circuits tailored to specific problems. ADAPT-VQE~\cite{grimsley2019adaptive} exemplifies this approach by iteratively building ansätze through greedy selection of operators that maximize energy reduction at each step. This adaptive strategy improves upon standard VQE by focusing computational resources on the most impactful circuit elements, rather than using a generic fixed structure. However, existing ADAPT-VQE implementations face a critical limitation: they either rely on full-space VQE evaluations or sacrifice expressivity through restrictive operator pools, treating ansatz design independently from the classical simulation efficiency constraints~\cite{shkolnikov2023avoiding, holmes2022connecting}.

When VQEs are implemented on classical simulators, cost function evaluation becomes a computational bottleneck~\cite{mccaskey2018validating}. Full statevector simulation scales exponentially as $\mathcal{O}(2^n)$~\cite{sim2019expressibility}, rendering large systems intractable. This limitation has motivated development of specialized classical simulators that exploit circuit structure to achieve polynomial-time cost function evaluation.

\subsection{Stabilizer Tensor Network Simulators}

Stabilizer Tensor Network (STN) simulators~\cite{masot2024stabilizer} combine the stabilizer formalism~\cite{gottesman1997stabilizer} with tensor network contraction to enable polynomial-time classical evaluation of quantum expectation values for Clifford-dominated circuits. Clifford gates, which are unitaries that map Pauli operators to Pauli operators, can be tracked classically in $\mathcal{O}(n^3)$ time via stabilizer tableaux. Non-Clifford gates are handled by a tensor network whose contraction cost scales with the number of such gates. Consequently, reducing the non-Clifford gate count directly lowers the simulation cost, which is the motivation for our Clifford rounding technique (Section~\ref{sec:implementation}).

\subsection{Qubit Reduction and Contextual Subspace Methods}
Ground states of molecular systems concentrate amplitude on a low-dimensional manifold, enabling qubit reduction by projecting the Hamiltonian onto a smaller subspace~\cite{cao2019quantum, mcclean2016theory}. Contextual subspace methods exploit this structure by identifying low-energy configurations and intelligently projecting the optimization problem onto carefully chosen subspaces $\mathcal{H}_\text{sub} \subseteq \mathcal{H}_\text{full}$ that are compatible with physical symmetries and electronic structure properties~\cite{khairy2020learning}, reducing the Hilbert space from $2^n$ to $2^{n'}$ dimensions where $n' \ll n$. Our pipeline uses Contextual Subspace~\cite{Weaving2025} for this step, though other reduction methods such as active space selection or entanglement forging could serve the same role.

\subsection{Sample-based Quantum Diagonalization}
SQD~\cite{Robledo_Moreno_2025}, which builds on the quantum-selected configuration interaction (QSCI) framework~\cite{kanno2023qsci}, uses a quantum circuit purely as a bitstring sampler: sampled configurations are used to construct a reduced configuration interaction subspace, which is then classically diagonalized to refine the ground-state energy. Because SQD performs its own classical eigen-solve, it does not require the VQE circuit to produce an accurate energy estimate---only bitstrings with sufficient overlap on the ground-state subspace. This property is central to our work: it means the preceding circuit can be aggressively simplified (fewer operators, Clifford-rounded parameters) without degrading the final SQD accuracy.

\begin{figure*}[t]
    \centering
    \includegraphics[width=0.8\linewidth]{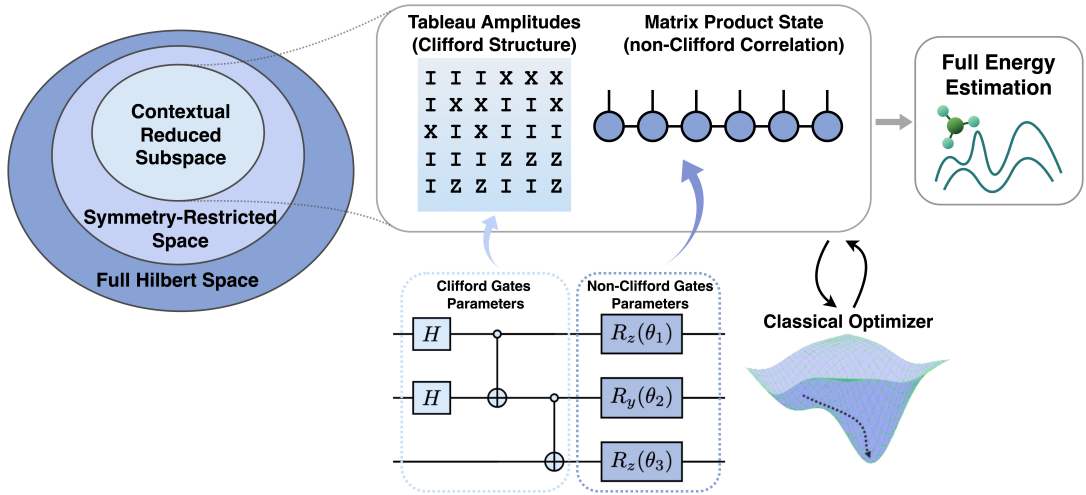}
    \caption{Overview of the proposed pipeline. The full Hilbert space is compressed into a contextual reduced subspace capturing the dominant low-energy structure of the molecule. Inside this subspace, the quantum state is expressed via a dual representation: a stabilizer-tableau component encoding Clifford-compatible structure and an MPS component capturing non-Clifford correlations. During variational optimization, Clifford gates act only on the tableau representation and non-Clifford gates act only on the MPS component, coordinated by a classical optimizer. The optimized reduced-space state is finally lifted back to the full orbital space using a precomputed inverse mapping, enabling accurate reconstruction of molecular observables and ground-state energies.}
    \label{fig:overview}
\end{figure*}

\section{Method}
\subsection{Core Insight}
Quantum chemistry simulation reveals a fundamental contradiction: the full electronic structure problem requires exploring exponentially large Hilbert spaces, yet near-term quantum processors cannot maintain coherence for the deep circuits necessary to explore such spaces. Classical simulation conversely can execute arbitrarily deep circuits but faces exponential overhead.

Our core insight is that SQD changes what the quantum circuit must achieve. In standalone VQE, the circuit must produce an accurate energy expectation value, demanding high expressivity and deep circuits. With SQD as a post-processing step, the circuit only needs to produce bitstrings that overlap sufficiently with the ground-state subspace which is a much weaker requirement. This gap between what standalone VQE demands and what SQD actually needs creates room for two circuit compression techniques:

\textbf{Gradient pruning} removes excitation operators with small gradient magnitude, reducing the number of ansatz parameters. \textbf{Clifford rounding} snaps the parameters of the least significant remaining operators to the nearest Clifford angle ($k\pi/2$), converting non-Clifford gates into classically trackable Clifford operations. Both techniques degrade standalone VQE accuracy but, as we demonstrate in Section~\ref{5.2}, leave SQD accuracy essentially unchanged.

The overall pipeline (Figure~\ref{fig:overview}) proceeds in three stages: (1) qubit reduction via contextual subspace projection, (2) compressed VQE with gradient pruning and Clifford rounding, and (3) SQD recovery.

\subsection{Stage 1: Qubit Reduction}
\label{sec:implementation}
The reduction stage applies qubit tapering followed by contextual-subspace projection. Beginning with a molecular Hamiltonian in Jordan-Wigner representation, we first exploit conservation laws to reduce the qubit count, yielding the tapered Hamiltonian $H_{\text{taper}}$ operating on $n_{\text{taper}} \ll n_{\text{full}}$ qubits with no approximation error.

Next, we construct stabilizer generators defining a low-energy subspace, projecting the tapered Hamiltonian onto this stabilizer eigenspace to yield the contextual-subspace Hamiltonian $H_{\text{cs}}$ on $n_{\text{cs}}$ qubits. We extract excitation operators by projecting UCCSD into the CS space, retaining only dominant Pauli terms to keep circuit depths manageable. We also precompute a mapping table relating CS bitstrings to the original Jordan-Wigner space, stored for use in Stage 3 recovery. All Stage 1 computations are performed once offline and cached.

\subsection{Stage 2: Compressed Variational Optimization}
The ansatz is parameterized as $|\psi(\theta)\rangle = \prod_{j=1}^{N} e^{-i\theta_j P_j} |{\mathrm{HF}}\rangle$, where $P_j$ are single-term Pauli excitations and $|{\mathrm{HF}}\rangle$ is the Hartree-Fock reference prepared with Clifford gates.

\textbf{Gradient Pruning}: We first compute gradients $\partial E / \partial \theta_j$ for all operators using a shallow reference circuit, evaluating each excitation operator independently at $\theta_j = 0$ with all other parameters fixed at zero. This single-term gradient evaluation is efficient: it requires $N$ STN evaluations rather than the $\mathcal{O}(N^2)$ evaluations that a full ADAPT-VQE gradient step would need, since we do not account for inter-operator interactions. We then select the top-ranked operators by gradient magnitude up to a specified fraction $f_g$ of the full pool, discarding operators whose gradients fall below the threshold. The discarded operators' parameters are fixed at zero, effectively removing those gates from the circuit entirely. This reduces parameter count without sacrificing expressivity for the dominant effects.

\textbf{Clifford Rounding}: After VQE convergence, we identify the $\lfloor f_c \cdot N_{\text{active}} \rfloor$ parameters closest to a Clifford angle $k\pi/2$ and snap them to that angle, where $f_c$ is the Clifford fraction controlling what proportion of active parameters are rounded. These gates become Clifford operations that the STN simulator handles at negligible cost, directly reducing both simulation time and hardware circuit depth. The remaining $N_{\text{active}} - \lfloor f_c \cdot N_{\text{active}} \rfloor$ parameters retain their optimized non-Clifford values, preserving expressivity where it matters most.

Concretely, the rounding uses a min-diff strategy: for each parameter $\theta_j$, we compute $\delta_j = \min_k |\theta_j - k\pi/2|$, the distance to the nearest Clifford angle. We sort parameters by $\delta_j$ in ascending order and round the $\lfloor f_c \cdot N_{\text{active}} \rfloor$ parameters with the smallest $\delta_j$. This ensures that we round the parameters that are already closest to a Clifford angle, minimizing the perturbation to the wavefunction. Parameters far from any Clifford angle, typically those encoding the strongest correlations, are left untouched. When $f_c = 0$ no parameters are rounded (fully non-Clifford); when $f_c = 1$ all parameters are rounded (fully Clifford). Note that Clifford rounding is applied after VQE convergence: the optimizer runs with full non-Clifford freedom, and rounding is a post-hoc compression step. This is important because it means the rounding does not affect the optimization landscape, only the final circuit used for SQD sampling.

The cost function is evaluated using the STN simulator, which exploits the Clifford-dominance of the circuit, computing the expectation value in polynomial time by tracking the stabilizer state. Fewer non-Clifford gates directly lower the STN contraction cost. Upon convergence, we sample bitstrings from the final state in the computational basis. These samples serve as input to Stage 3.

\subsection{Stage 3: SQD Recovery}
Bitstrings sampled from the VQE state are lifted to the original Jordan-Wigner space using the precomputed mapping table, accounting for weighted contributions from the eliminated degrees of freedom. The lifted bitstrings are then passed to SQD's fermion-solver, which builds a selected-CI subspace from the sampled determinants and solves it via classical diagonalization to refine energy estimates.

In each iteration, SQD generates fresh samples, constructs configuration space approximations to the full Hamiltonian using sampled determinants, and solves these approximations exactly via classical diagonalization. Over iterations, the algorithm converges toward the ground state. Specifically, each SQD iteration proceeds as follows: (i)~bitstrings are postselected to the correct electron number in each spin sector; (ii)~the postselected bitstrings are subsampled into multiple batches to reduce variance; (iii)~for each batch, a configuration interaction (CI) matrix is constructed from the PySCF one- and two-electron integrals restricted to the sampled determinants; (iv)~a Davidson eigensolver diagonalizes each CI matrix to yield a batch energy estimate; (v)~the per-batch occupancy profiles are averaged and fed back to a configuration recovery step that biases the next iteration's sampling toward the most occupied orbitals. The best energy across all batches and iterations is reported as the final SQD estimate.

\section{Evaluation}
\subsection{Experiment Setup} \label{5.1}
\paragraph{Problem Instances} We study a set of 21 small molecules and ions in the STO-3G minimal basis, including H$_2$O, N$_2$, HCl, and map them to qubits via the Jordan–Wigner transformation. For each system, we generate its Pauli-sum Hamiltonian, record qubit and electron counts, the Hartree–Fock reference state vector, and other molecular attributes from PySCF~\cite{10.1063/5.0006074}. Across the set, the full JW register spans 6–20 qubits, depending on the system.

\paragraph{Contextual Subspace} To reduce problem size while preserving ground‑state accuracy, we apply qubit tapering with the HF reference and then project the tapered Hamiltonian into a contextual subspace~\cite{Weaving2025}. For each molecule we sweep the subspace size and select the smallest number of contextual qubits whose subspace ground energy is within approximately 1.6 mHa of full configuration interaction (FCI). We also compute a mapping from the contextual subspace back to the full Jordan–Wigner basis with associated weights, so that sampled contextual configurations can be consistently distributed over the full basis during the SQD step. Within the contextual subspace we initialize the HF determinant and build a single‑term, UCCSD‑inspired excitation pool by projecting the tapered UCC operator and retaining the dominant Pauli term from each composite operator.

\paragraph{VQE setting} For the VQE used in our experiments, we choose COBYLA optimizer with initial parameters set to zero and reports electronic energies including nuclear repulsion. For the simulation estimator, we adopt and modify the implementation from~\cite{masot2024stabilizer} to make our own STN estimator. For hardware experiments, circuits are transpiled to the \texttt{ibm\_rensselaer} instruction set at optimization level 3 and sampled with 100k shots via IBM Qiskit SamplerV2.

\paragraph{SQD setting} SQD post‑processing proceeds by sampling bitstrings from the optimized contextual ansatz. On simulation, we sample from the exact statevector distribution; on real hardware, we measure with IBM Qiskit SamplerV2. Sampled contextual bitstrings are lifted to JW via the precomputed CS to JW map, postselected to the correct electron numbers. We then perform a subspace eigen‑solve per batch, which builds a selected‑CI subspace using the PySCF one‑ and two‑electron integrals and solves it with a Davidson routine. Unless noted otherwise, SQD runs for 5 outer iterations; we cap Davidson iterations at 2000 and use a fixed random seed 42.

\paragraph{Compression settings} We evaluate gradient pruning at fractions $f_g \in \{1.00, 0.75, 0.50, 0.25\}$ and Clifford rounding at fractions $f_c \in \{1.00, 0.75, 0.50, 0.25, 0.00\}$, forming a $4 \times 5$ ablation grid. On hardware, we test four representative configurations: baseline ($f_g{=}1, f_c{=}1$), gradient only ($f_g{=}0.5$), Clifford only ($f_c{=}0.5$), and the combination of gradient and clifford method ($f_g{=}0.5, f_c{=}0.5$).

\subsection{Experiment Results} \label{5.2}

\paragraph{Baseline Performance}

\begin{figure}[t]
    \centering
    \includegraphics[width=.9\linewidth]{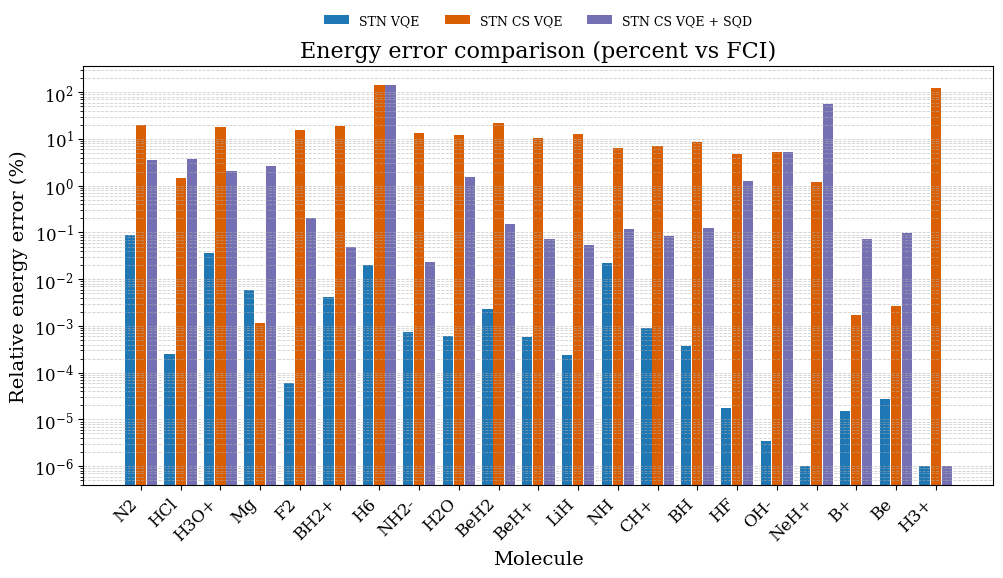}
    \caption{Relative energy error versus FCI comparison under full parameters mode}
    \label{fig:energy_error_full}
    \vspace{-.3cm}
\end{figure}

We first establish the baseline by comparing three methods without any circuit compression: standard VQE on the full Hamiltonian, CS-VQE (qubit-reduced, no SQD), and our full pipeline CS-VQE+SQD. Figure~\ref{fig:energy_error_full} shows the relative energy error versus FCI. The baseline VQE consistently achieves very low error, around $10^{-3}\%$. CS-VQE alone shows much higher error in many cases, around $1 - 10\%$ or worse, because truncating to a small contextual subspace sacrifices expressive power. When SQD is added, the error in most molecules falls back into the $\leq 10^{-2}\%$ region. This confirms that SQD effectively recovers the accuracy lost through subspace truncation---the property we exploit for circuit compression.

\begin{figure}[t]
    \centering
    \includegraphics[width=.95\linewidth]{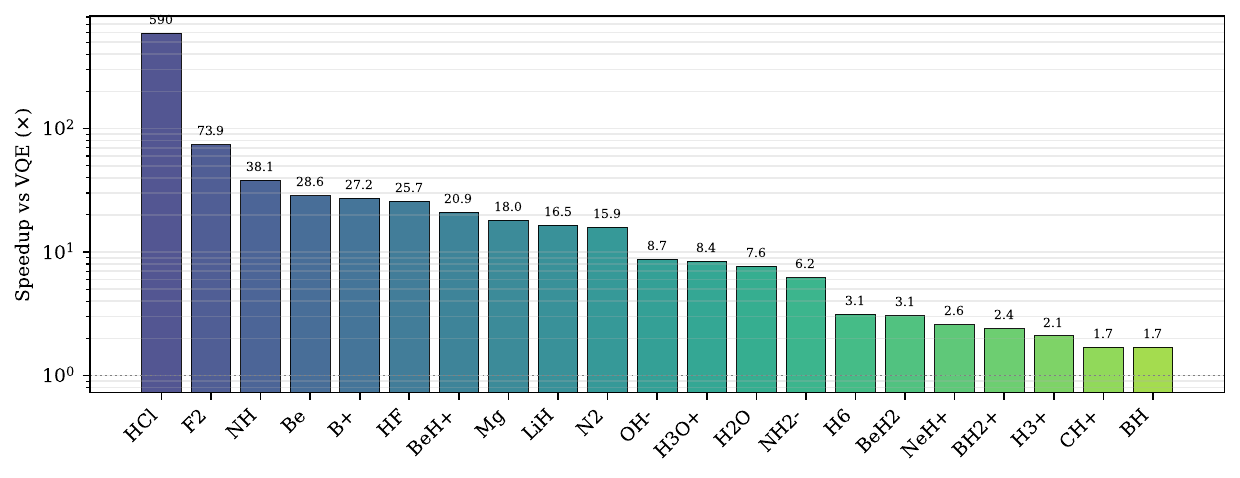}
    \caption{Runtime speedup of CS-VQE+SQD vs.\ standard VQE across 21 molecules (uncompressed baseline, $f_g{=}1, f_c{=}1$). Molecules are sorted by descending speedup. HCl achieves $590\times$ due to aggressive qubit reduction (20$\to$3 qubits); molecules with large contextual subspaces (BH, CH$^+$) show modest gains.}
    \label{fig:runtime_speedup}
\end{figure}

Figure~\ref{fig:runtime_speedup} shows the runtime speedup of our pipeline (CS-VQE+SQD) relative to the standard VQE baseline across all 21 molecules. The speedups vary substantially: HCl achieves $590\times$ with 20 JW qubits reduced to 3 CS qubits, F$_2$ achieves $74\times$, NH $38\times$, and Be/B$^+$ around $28\times$. At the lower end, BH and CH$^+$ achieve only $1.7\times$ because their contextual subspaces remain large: 8 CS qubits out of 12 JW qubits. Of the 21 molecules, 15 achieve SQD error below the 1.6\,mHa chemical accuracy threshold. The outliers: N$_2$ (33.8\,mHa, 14 CS qubits), Mg (17.5\,mHa), H$_3$O$^+$ (16.7\,mHa) require large contextual subspaces, limiting the compression benefit. These molecules are included in all subsequent ablation results to ensure our compression techniques are evaluated on both easy and hard cases.

\paragraph{Ablation Study: Gradient Pruning $\times$ Clifford Rounding}

\begin{figure*}[t]
    \centering
    \includegraphics[width=.85\textwidth]{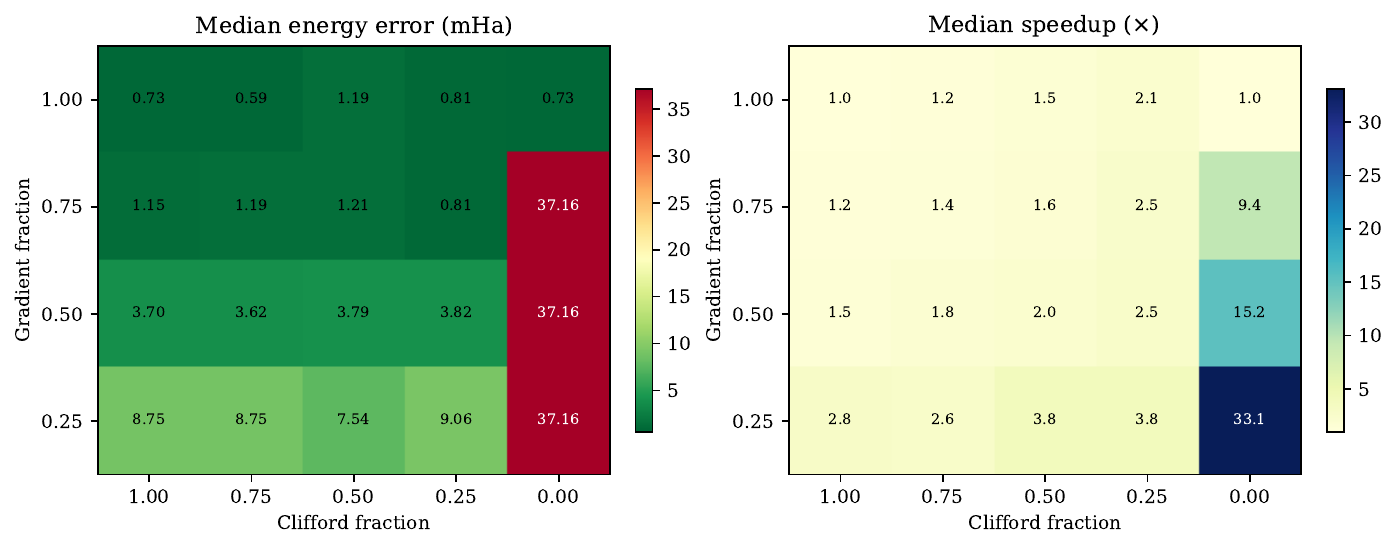}
    \caption{Ablation heatmap across 21 molecules. Left: median SQD energy error (mHa). Right: median simulation speedup ($\times$). Rows vary gradient fraction $f_g$; columns vary Clifford fraction $f_c$. Error remains within chemical accuracy (1.6\,mHa, bold in Table~\ref{tab:ablation}) for the top-left region; speedup increases toward the bottom-right. The $f_c = 0.00$ column shows a phase transition: accuracy collapses when no non-Clifford gates remain, but speedup peaks.}
    \label{fig:ablation_heatmap}
\end{figure*}

The central experiment is a systematic ablation over both compression axes. Figure~\ref{fig:ablation_heatmap} and Table~\ref{tab:ablation} show the median energy error and speedup across all 21 molecules for each $(f_g, f_c)$ combination.

\begin{table}[t]
\centering
\caption{Median SQD energy error (mHa) across 21 molecules. Bold entries are within chemical accuracy (1.6\,mHa). Rows: gradient fraction $f_g$. Columns: Clifford fraction $f_c$.}
\label{tab:ablation}
\scriptsize
\begin{tabular}{lrrrrr}
\toprule
$f_g \downarrow$ / $f_c \rightarrow$ & 1.00 & 0.75 & 0.50 & 0.25 & 0.00 \\
\midrule
  1.00 & \textbf{0.73} & \textbf{0.59} & \textbf{1.19} & \textbf{0.81} & \textbf{0.73} \\
  0.75 & \textbf{1.15} & \textbf{1.19} & \textbf{1.21} & \textbf{0.81} & 37.16 \\
  0.50 & 3.70 & 3.62 & 3.79 & 3.82 & 37.16 \\
  0.25 & 8.75 & 8.75 & 7.54 & 9.06 & 37.16 \\
\bottomrule
\end{tabular}
\end{table}

Several findings emerge. First, Clifford rounding at $f_c \geq 0.25$ has almost no effect on accuracy for a given $f_g$: error stays flat across columns until $f_c = 0.00$ where it jumps sharply. This confirms that most VQE parameters are already near Clifford angles and rounding them is nearly free. Second, gradient pruning is the dominant source of error increase: reducing $f_g$ from 1.00 to 0.25 raises median error from 0.73 to 8.75\,mHa, regardless of $f_c$. Third, the sweet spot for chemical accuracy is $f_g \geq 0.75$ with any $f_c \geq 0.25$, achieving a median simulation speedup of $1.2$--$2.5\times$ from Clifford rounding alone and up to $9.4\times$ when combined with gradient pruning.

Table~\ref{tab:ablation_speedup} shows the corresponding median simulation speedups. The two compression axes contribute multiplicatively: gradient pruning at $f_g = 0.25$ alone gives $2.8\times$, Clifford rounding at $f_c = 0.25$ alone gives $2.1\times$, and combining them yields $3.8\times$. The $f_c = 0.00$ column achieves the largest speedups ($9$--$33\times$) because the STN simulator handles fully Clifford circuits at negligible cost, but as Table~\ref{tab:ablation} shows, this comes at the price of accuracy. The practical operating point is therefore $f_c \in [0.25, 0.50]$, which captures most of the speedup benefit while staying well within chemical accuracy.

\begin{table}[t]
\centering
\caption{Median simulation speedup ($\times$) vs baseline across 21 molecules. Rows: gradient fraction $f_g$. Columns: Clifford fraction $f_c$.}
\label{tab:ablation_speedup}
\scriptsize
\begin{tabular}{lrrrrr}
\toprule
$f_g \downarrow$ / $f_c \rightarrow$ & 1.00 & 0.75 & 0.50 & 0.25 & 0.00 \\
\midrule
  1.00 & 1.00$\times$ & 1.17$\times$ & 1.52$\times$ & 2.06$\times$ & 1.01$\times$ \\
  0.75 & 1.19$\times$ & 1.37$\times$ & 1.63$\times$ & 2.47$\times$ & 9.37$\times$ \\
  0.50 & 1.49$\times$ & 1.79$\times$ & 2.04$\times$ & 2.51$\times$ & 15.22$\times$ \\
  0.25 & 2.81$\times$ & 2.57$\times$ & 3.78$\times$ & 3.80$\times$ & 33.09$\times$ \\
\bottomrule
\end{tabular}
\end{table}

We note that the ablation grid reveals molecule-dependent behavior. Molecules with few CS qubits and few excitation operators (e.g., Be, B$^+$, HCl with 3 CS qubits and 6--16 operators) are robust to aggressive compression, maintaining chemical accuracy even at $f_g = 0.25$, because SQD has sufficient bitstring diversity from their small Hilbert spaces. Conversely, molecules with large contextual subspaces (e.g., N$_2$ with 14 CS qubits, H$_3$O$^+$ with 10 CS qubits) are more sensitive to pruning because the ground state requires many excitation operators to achieve adequate overlap. For these systems, $f_g \geq 0.75$ is recommended to maintain accuracy, while Clifford rounding remains safe at any $f_c \geq 0.25$.

\paragraph{Individual Sweep Analysis}
Sweeping each axis in isolation confirms these trends. Reducing $f_g$ from 1.00 to 0.50 keeps the median error below chemical accuracy while the median speedup grows to $1.5\times$; at $f_g = 0.25$ the speedup reaches $2.8\times$ but error rises above the threshold for several molecules, with the steepest degradation seen in molecules that rely on many excitation operators. Clifford rounding shows error essentially flat from $f_c = 1.00$ to $0.25$ (variance across molecules stays low in this range) while speedup climbs to $2.1\times$; only at $f_c = 0.00$, where the circuit loses all non-Clifford expressivity, does accuracy degrade sharply. Both axes thus offer a favorable accuracy--speedup tradeoff, with Clifford rounding the cheaper of the two in accuracy cost.

An interesting anomaly appears at $f_c = 0.00$ in Table~\ref{tab:ablation}: the median error at $(f_g{=}1.00,\, f_c{=}0.00)$ is 0.73\,mHa, identical to the uncompressed baseline $(f_g{=}1.00,\, f_c{=}1.00)$. Yet it jumps to 37.16\,mHa for $f_g \leq 0.75$. This reveals an interaction between the two compression axes. When all UCCSD operators are present ($f_g{=}1.00$), the fully Clifford circuit still distributes bitstring probability across configurations that SQD can exploit. But when operators are also pruned, the combined loss of both operator diversity and non-Clifford expressivity pushes the circuit’s output distribution too far from the ground-state subspace, and SQD can no longer recover. This finding underscores that the two techniques are not independently safe at their extremes: practitioners should avoid combining aggressive pruning with full Clifford rounding.

\paragraph{Hardware Validation}

\begin{table*}[t]
\centering
\caption{Hardware results on \texttt{ibm\_rensselaer}. Transpiled circuit depth and SQD energy error across four configurations. All runs use 100k shots.}
\label{tab:hardware}
\scriptsize
\begin{tabular}{lrrrrrrrr}
\toprule
 & \multicolumn{4}{c}{Transpiled Depth} & \multicolumn{4}{c}{SQD Error (mHa)} \\
\cmidrule(lr){2-5} \cmidrule(lr){6-9}
Mol. & Baseline & Grad 0.50 & Cliff 0.50 & Combined & Baseline & Grad 0.50 & Cliff 0.50 & Combined \\
\midrule
  Be & 31 & 33 & 35 & 33 & 0.33 & 0.33 & 0.33 & 0.33 \\
  B$^+$ & 31 & 21 & 35 & 21 & 0.38 & 0.38 & 0.38 & 0.38 \\
  LiH & 711 & 299 & 468 & 252 & 1.30 & 1.30 & 1.30 & 1.30 \\
  BeH$^+$ & 306 & 117 & 249 & 116 & 0.32 & 0.32 & 0.32 & 0.32 \\
  HCl & 161 & 100 & 142 & 58 & 0.54 & 0.54 & 0.54 & 0.54 \\
  H$_2$O & 4437 & 2227 & 3378 & 1690 & 1.14 & 1.14 & 1.14 & 1.14 \\
\midrule
  Median red. & 1.0$\times$ & 1.8$\times$ & 1.2$\times$ & 2.6$\times$ & & & & \\
\bottomrule
\end{tabular}
\end{table*}

We validate on \texttt{ibm\_rensselaer} with 6 molecules spanning 3--6 CS qubits. Table~\ref{tab:hardware} presents the results. The most striking finding is that \textbf{SQD energy error is identical across all four configurations for every molecule}. Gradient pruning, Clifford rounding, and their combination produce exactly the same SQD accuracy as the uncompressed baseline. This is the direct experimental confirmation of our core insight: SQD’s tolerance to circuit imperfection makes these compressions free in terms of accuracy.

\begin{figure}[t]
    \centering
    \includegraphics[width=.9\linewidth]{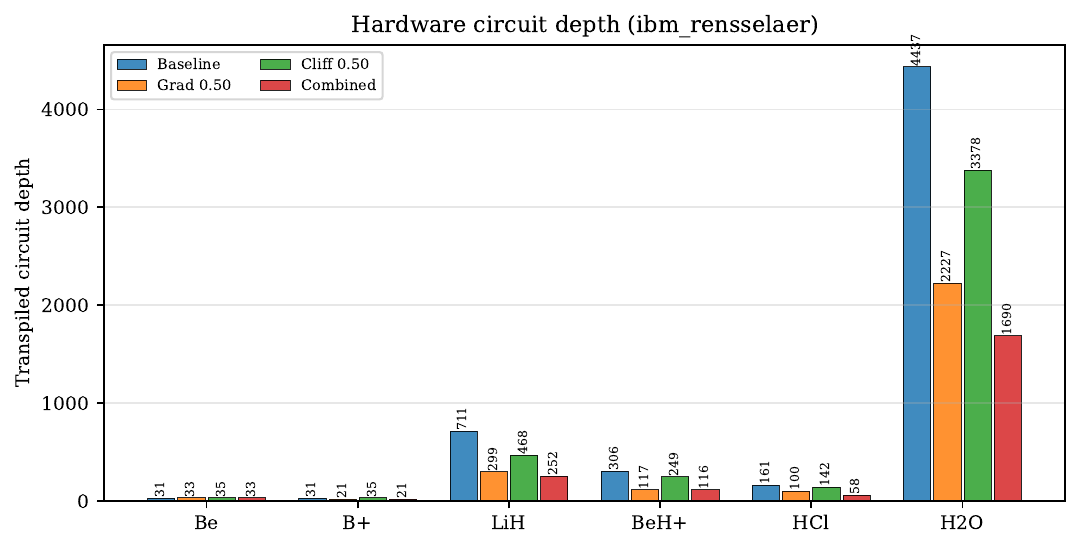}
    \caption{Absolute transpiled circuit depth on \texttt{ibm\_rensselaer} across four configurations. The depth reduction scales with molecule size: H$_2$O (168 parameters) drops from 4437 to 1690; small molecules (Be, B$^+$) are already shallow and show no change.}
    \label{fig:hw_depth_abs}
\end{figure}

Meanwhile, the transpiled circuit depth drops substantially (Figure~\ref{fig:hw_depth_abs}). For molecules with more than a few parameters, the combined configuration reduces depth by $2.4$--$2.8\times$: LiH drops from 711 to 252, BeH$^+$ from 306 to 116, HCl from 161 to 58, and H$_2$O from 4437 to 1690. Gradient pruning contributes the larger share (median $1.8\times$), while Clifford rounding adds an additional $1.2\times$ on top. Small molecules (Be, B$^+$ with 6 parameters and depth $\sim$30) show no benefit, as their circuits are already shallow. The depth reduction is most significant for H$_2$O (168 parameters), where the combined approach removes 2747 layers of circuit depth, directly translating to reduced decoherence and gate errors on hardware.

The invariance of SQD accuracy across all four hardware configurations arises because all four circuits, despite varying compression levels, still produce bitstrings with high overlap on the ground-state subspace: the CS qubit counts here are small enough that even a significantly perturbed circuit distributes probability over the correct region, letting SQD's eigensolve extract the same optimal energy from every sample set. This suggests SQD-based pipelines are particularly well-suited to molecules where qubit reduction yields a small contextual space, since the needed bitstring diversity is easily achievable even with compressed circuits.

From a hardware-cost perspective, the depth reduction directly impacts circuit fidelity: on \texttt{ibm\_rensselaer} (Eagle r3, 127 qubits) with typical two-qubit gate error rates of $\sim 10^{-2}$, a circuit of depth 4437 (H$_2$O baseline) accumulates substantially more noise than one of depth 1690 (combined). Both produce identical SQD energy estimates in our experiments, but the compressed circuits provide additional headroom for noisier backends or larger molecules where accumulated errors could eventually degrade bitstring quality below SQD's tolerance threshold.

\section{Related Work}

\textbf{Clifford-assisted VQA.}
CAFQA~\cite{ravi2022cafqa} restricts ansätze to Clifford gates for classical simulability via Bayesian optimization, and \cite{10.1145/3622781.3674178} extends this to mixed Clifford/non-Clifford circuits. These methods use Clifford structure to make optimization cheaper; our Clifford rounding instead exploits it to compress an already-optimized circuit for SQD sampling. Unlike CAFQA, our approach retains non-Clifford parameters where they matter most and relies on SQD to compensate where they are rounded away.

\textbf{Adaptive ansatz construction.}
ADAPT-VQE~\cite{grimsley2019adaptive} iteratively grows the ansatz by selecting operators that maximize energy reduction. Our gradient pruning is related in spirit, also using gradient information to select operators, but ADAPT-VQE builds the ansatz from scratch while we prune a pre-defined UCCSD pool, and it targets VQE energy directly while our pruning targets a weaker objective that permits more aggressive reduction.

\textbf{VQE parameter optimization.}
SOAP~\cite{li2024efficient} reduces energy evaluations via sequential parabolic approximation, and NAPA~\cite{liang2024napa} optimizes at the pulse level; both improve efficiency in reaching a VQE solution but do not address what happens after. Our work is complementary: any VQE optimizer can be used in Stage~2, with compression applied post-convergence.

\textbf{Subspace methods and SQD.}
The sample-and-diagonalize framework was first proposed as QSCI~\cite{kanno2023qsci}; SQD~\cite{Robledo_Moreno_2025} extends it with configuration recovery and large-scale demonstrations on nitrogen and iron--sulfur clusters using fixed, carefully optimized circuits. ADAPT-QSCI~\cite{nakagawa2024adaptqsci} adaptively grows the input circuit through iterative operator selection, sharing our goal of a compact sampling circuit; we instead prune a fixed UCCSD pool post hoc in a qubit-reduced space. Subsequent work has used SQD with standard VQE circuits as an accuracy booster for a given circuit quality, but no prior work has asked the inverse question: \emph{given that SQD will post-process the output, how much can the preceding circuit be degraded?} Our contribution answers this directly, showing that circuit quality can be substantially degraded via pruning and rounding without affecting SQD's output, reframing SQD not merely as an accuracy refinement tool but as an enabler of circuit compression.

\textbf{Contextual subspace methods.}
Contextual subspace VQE~\cite{Weaving2025} reduces qubit count by projecting onto stabilizer-defined subspaces; in our pipeline it provides the qubit reduction of Stage~1 but is not the main contribution, and any qubit-reduction method could serve the same role (Section~II-C). Our novelty lies in the Stage~2 circuit compression and the demonstration that SQD tolerates it.

\textbf{Circuit compilation and depth reduction.}
Transpiler-level optimizations~\cite{ding2020systematic} reduce circuit depth through gate cancellation, routing, and scheduling, independently of the algorithm. Our compression is complementary: gradient pruning and Clifford rounding simplify the logical circuit before transpilation, so the $2.6\times$ median depth reduction we observe (Table~\ref{tab:hardware}) stacks on top of whatever the transpiler achieves at optimization level~3. Combining algorithm- and transpiler-level compression is a promising direction for maximizing circuit fidelity on near-term hardware.

\section{Conclusion}
We showed that SQD's tolerance to circuit imperfection creates a previously unexploited opportunity for circuit compression. By introducing gradient pruning and Clifford rounding, we reduce transpiled circuit depth by up to $2.8\times$ on IBM hardware while SQD preserves identical energy accuracy. A systematic ablation across 21 molecules characterizes the two-dimensional tradeoff: Clifford rounding is nearly free in accuracy cost, gradient pruning dominates the error budget, and their combination yields the largest depth savings, a finding confirmed by hardware experiments on 6 molecules where all configurations produce the same energy error as the uncompressed baseline. These results suggest that future SQD-based pipelines should deliberately design for shallow, imprecise circuits rather than optimizing VQE accuracy, since SQD will recover the accuracy regardless.

\textbf{Limitations and future work.} Our evaluation uses the STO-3G minimal basis with 3--14 CS qubits that remain classically simulable; extending to larger basis sets beyond 20 CS qubits, generalizing the compression principle beyond UCCSD to other ansätze, co-optimizing parameters with a rounding-aware objective, and combining with error mitigation are promising directions for future work.
\bibliographystyle{ACM-Reference-Format}
\bibliography{sample-base}
\end{document}